-------------------------------------------------------------------------------
\documentstyle[12pt]{article}
\begin{document}
\sloppy
\begin{center}
{\Large\bf Quantisation of second class systems in the\\
Batalin-Tyutin formalism}\\[.5cm]
N. Banerjee\footnote{e-mail address: shila@saha.ernet.in}\\[.5cm]
 Saha Institute of Nuclear Physics, 1/AF Bidhannagar\\
Calcutta 700064, India\\[.5cm]
 R. Banerjee \\[.5cm]
 S. N. Bose National center for Basic Sciences\\
DB-17, Salt Lake, Calcutta 700064, India\\[.5cm]
 Subir Ghosh\\[.5cm]
 Gobardanga Hindu College, North 24 Parganas\\
 West Bengal, India\\[1cm]
\end{center}
\begin{abstract}
We review the Batalin-Tyutin approach of quantising second class systems  which
consists in enlarging the phase space to convert such systems into first class.
The
quantisation  of  first  class  systems,  it  may be mentioned, is already well
founded. We show how the usual analysis of Batalin-Tyutin may  be  generalised,
particularly  if  one  is  dealing with nonabelian theories. In order to gain a
deeper insight into the formalism we have considered two specific  examples  of
second class theories-- the massive Maxwell theory (Proca model) and its
nonabelian
extension.  The  first  class  constraints  and  the involutive Hamiltonian are
explicitly constructed. The connection of our  Hamiltonian  approach  with  the
usual  Lagrangian  formalism  is  elucidated. For the Proca model we reveal the
importance of a boundary term which plays a significant role in establishing an
exact
identification of the extra fields in  the  Batalin-Tyutin  approach  with  the
St\"uckelberg  scalar. Some comments are also made concerning the corresponding
identification in the nonabelian example.
\end{abstract}
\newpage
\section {\bf Introduction}

Canonical  quantisation  of  systems with first class constraints was
formulated
along general lines by  Dirac  \cite{r1}.  The  corresponding  analysis  in
the  path
integral  approach  was  initiated   by  Faddeev  \cite{r2} for gauge theories.
It was
extended by Fradkin and collaborators \cite{r3,r4}  within  the  broader
framework  of
preserving  Becchi-Rouet-Stora-Tyutin (BRST) \cite{r5} invariance. The
quantisation of
systems with second class constraints, on the contrary, poses problems. In
this
case  it  is  necessary to replace the canonical Poisson brackets (P.B) by
their
corresponding Dirac brackets (D.B). The  conversion  of  the  D.B's  to
quantum
commutators  is,  in  general,  plagued  with  severe  factor ordering
problems.
Moreover, the abstraction of  the  canonically  conjugate  variables  is
highly
nontrivial. Consequently the quantisation of second class systems, either in
the
canonical or in  the  path  integral  formalisms,  is  problematic.  It  may
be
mentioned  that ther is a factor ordering problem of a different nature in
first
class  systems.  There  are  three  varities  of  this  problem.  First,   the
constraints  which  close  under  P.B  may  do  so  with  coefficients which
are
functions of the canonical  variables  rather  than  just  structure
constants.
Second,  the P.B of the constraints with the Hamiltonian may yield a
combination
of constraints with phase space coefficients. Finally, the  Poisson  algebra
of
the  constraints  with  any  physical  variable  must  yield  a  combination
of
constraints  which  may  involve  structure  functions  instead   of
structure
constants.  In  all  these cases a suitable factor ordering has to be found
such
that the structure  function  operators  {\it  precede}  the  correctly
ordered
constraints   \cite{r1,r6}.  It  is  only  then  possibly  to  develop  a
consistent
quantisation program. This has been done by Kuchar \cite{r6} in  a  covariant
manner.
The  factor  ordering  concerning  the  interpretation  of  the D.B as a
quantum
commutator remains an open issue.

In view of the above discussion it becomes natural to formulate the
quantisation
of second class systems without invoking D.B. A possible way would be  to
embed
such  a  system in an extended space so that it gets converted into first
class.
One can then apply the well established machinery  \cite{r3,r4,r5}  for
quantising  first
class  systems. This philosophy has been  recently  adopted  by
Batalin-Fradkin
\cite{r7}
and Batalin-Tyutin \cite{r8}. The phase space is extended by introducing new
variables
which  transform  the  original  second  class  system  into  first class. It
is
worthwhile to mention that this idea is a logical  follow  up  of  the
original
notion  of  St\"uckelberg  \cite{r9}  who  extended the configuration space to
convert
second class theories into first class. The approach of St\"uckelberg \cite{r9}
is  in
the  Lagrangian  formulation  which  should  be  contrasted with the
Hamiltonian
formulation of Batalin-Fradkin \cite{r7} or Batalin-Tyutin \cite{r8}. A similar
(Lagrangian)
functional   integral   approach   has   been  used  by  Faddeev  and
Satashvili
\cite{r10} to
introduce Wess-Zumino scalars \cite{r11} to interpret anomalous gauge theories
as true
({\it  i.e.}  first  class)  gauge systems. Recently there have been
suggestions
\cite{r12,r13} that the extra fields introduced in the  Hamiltonian  formalism
may  be
identified  with  the  St\"uckelberg  scalar  or the Wess-Zumino fields. We
will
return to this point later.

The purpose of  this  paper  is  to  make  a  thorough  investigation  into
the
Hamiltonian  formulation  of  Batalin-Tyutin \cite{r8} to convert second class
systems
into first class, by referring to two examples-- the abelian (Proca) model and
its
nonabelian extension. In the course of this analysis the precise connection
with
the  Lagrangian  formulation  as  well as the identification of the extra
fields
with the St\"uckelberg \cite{r9} scalar will also be elucidated.

In section 2 we first briefly review the analysis of  Batalin-Tyutin
\cite{r8}.  This
will  serve a twofold purpose; to set up the notation and familiarise the
reader
with the basics of the formalism. This formalism \cite{r8}  is  ideal  for
discussing
abelian (second class) systems. To appreciate this point it may be recalled
that,
for abelian first class systems, the usual algebra of constraints among
themselves  and
with  the  Hamiltonian  is strongly involutive. The method developed in
\cite{r8} also
yields a strongly involutive algebra for first  class  systems.  For
nonabelian
(first  class)  theories,  on  the  other  hand,  it  is  well  known  that
the
corresponding algebra is only weakly involutive. We, therefore,  generalise
the
approach  of  \cite{r8}  to include this possibility. This has been discussed
in great
details. Its utility is made transparent when we actually convert  a
nonabelian
second class theory into first class. We are able to reproduce the algebra
which
occurs in usual nonabelian first class theories. If the standard  proceedure
of
\cite{r8}  were  adopted we would obviously fail to generate this algebra. This
will
also have some implications in connecting the  Hamiltonian  formalism  with
the
Lagrangian version.

Section  3  is  devoted to an application of the ideas discussed in section 2
to
the abelian Proca model \cite{r9}. The original second  class  system  is
transformed
into  first  class by initially converting the second class constraints to
first
class and then changing the Hamiltonian into the corresponding involutive
form.
The  phase  space  partition function is constructed and explicitly evaluated
in
two special gauges. In the unitary gauge \cite{r7,r8} which corresponds to
taking  the
initial  second  class  constraints as the gauge fixing conditions, the
original
theory is reproduced. Nontrivial consequences are obtained in the
Faddeev-Popov
\cite{r14}  like  gauges  ({\it i.e.} gauges not involving the momenta). It
leads to a
Lagrangian embedding which reveals the first class nature  of  the  theory.
The
connection  with  the  conventional  St\"uckelberg  \cite{r9}  Lagrangian
mechanism is
established. It is shown how a boundary  term plays a crucial role in  making
a
one-to-one  correspondence  between the extra field in the Hamiltonian
formalism
with the St\"uckelberg \cite{r9} scalar.

In section 4 we introduce the nonabelian version of the Proca  model.  We
first
point  out  a  flaw  in  the conventional Dirac analysis performed by
Senjanovic
\cite{r15}. The correct D.B's among the canonical variables are worked out.
These  are
found  to  be  field  dependent.  We  next exploitthe generalised version of
the
Batalin-Tyutin \cite{r8} approach developed in section 2 to  convert  thye
nonabelian
second  class  system  into first class. It is interesting to observe that
while
one of the (first class) constraints  has a closed form,  the  other
constraint
and  the  involutive Hamiltonian are expressed by a power series in terms of
the
new fields and do not have any closed  expressions.  Specialising  to  the
case
where  the  gauge  group  is $SU(2)$, we are able to reproduce the gauge
algebra
given in  the  conventional  Lagrangian  formulation  \cite{r16} obtained by
the
nonabelian  St\"uckelberg  mechanism.  Although  the identification of the
extra
fields in the Hamiltonian formulation with the St\"ckelberg  fields  (which
now
appear  as the Euler angles \cite{r16}) in the Lagrangian approach is not so
direct as
in the abelian example,  nevertheless  there  is  no  conceptual  difficulty
in
understanding this correspondence.
Finally  we  show  that  contrary  to  claims  in the literature \cite{r17},
the
nonabelian extension of the usual St\"uckelberg  \cite{r9}  mechanism  fails
to
convert the nonabelian second class theory into first class.

Our concluding observations are given in section 5.

\vspace{1cm}
\section {\bf General formalism}

In  this section we first review the abelian conversion of a second class
system
into first class as developed by Batalin and  Tyutin (BT) \cite{r8}.  We  then
discuss  a
natural  extension  of  this approach which is suitable for analysing
nonabelian
models. Simultaneously our notations and conventions will also be specified.

Let us assume that the canonical variables, $(\phi_i(x),\pi^i(x))$ with $|i|=n$
and a Grassman parity $\epsilon(\phi_i) = \epsilon(\pi^i) = \epsilon_i$ define
the
initial phase space of a dynamical system. We further suppose that this system
contains a set of linearly independent bosonic second class constraints
$\Theta_\alpha$
which are some functions of the original canonical variables.
\begin{equation}
\Theta_\alpha = \Theta_\alpha (\phi,\pi),~ \epsilon_\alpha = 0,~ |\alpha| =
m<2n \label{eq:2.1}
\end{equation}
so that the matrix,
\begin{equation}
\Delta_{\alpha\beta}            (x,y)           =
\left\{\Theta_\alpha
(x),\Theta_\beta(y)\right\}\label{eq:2.2}
\end{equation}
has a nonvanishing determinant. The inclusion of other constraints ({\it i.e.}
fermion
or first class) is a matter of technical detail and poses no problems in
developing
the formalism.

We now convert the second class system into first class. The initial step is to
obtain the first class constraints starting from (\ref{eq:2.1}). Foliowing the
general philosophy,
new dynamical fields,
\begin{equation}
\Phi^\alpha,~~\epsilon_\alpha = 0,~~|\alpha| = m\label{eq:2.3}
\end{equation}
with the basic P.Bs,
\begin{equation}
\left\{\Phi^\alpha(x),~~\Phi^\beta(y)\right\} ~=
\omega^{\alpha\beta}(x,y)\label{eq:2.4}
\end{equation}
where $\omega$ is an invertible field independent antisymmetric matrix,
\begin{equation}
\omega^{\alpha\beta}(x,y) = -\omega^{\beta\alpha}(y,x),
{}~~\epsilon(\omega^{\alpha\beta})
= \epsilon_\alpha+\epsilon_\beta,\label{eq:2.5}\end{equation}
are introduced to extend the original phase space,
\begin{equation}
(\phi,\pi)\oplus (\Phi).\label{eq:2.6}
\end{equation}

The  new  first  class  constraints  of  the  system in the extended phase
space
(\ref{eq:2.6})
are denoted by $\tau_\alpha$:
\begin{equation}
\tau_\alpha = \tau_\alpha (\phi, \pi, \Phi);~~\epsilon(\tau_\alpha) =
\epsilon_\alpha,
{}~|\alpha| = m.\label{eq:2.7}
\end{equation}
Then the abelian conversion of BT \cite{r8} implies that these constraints are
strongly
involutive,
\begin{equation}
\left\{\tau_\alpha(x)~,~\tau_\beta(y)\right\} = 0\label{eq:2.8}
\end{equation}
subject to the boundary condition,
\begin{equation}
\tau_\alpha(\phi,\pi,0) = \Theta_\alpha(\phi, \pi)\label{eq:2.9}
\end{equation}
where the right hand side of (\ref{eq:2.9}) is just the original set of second
class constraints (\ref{eq:2.1}).

In ref. \cite{r8} it has been shown that it is possible to explicitly construct
the solution to the
algebraic problem (\ref{eq:2.8}) and (\ref{eq:2.9}) in the form of a power
series expansion,
\begin{equation}
\tau_\alpha (\phi, \pi, \Phi) = \sum_{n=0}^\infty \tau_\alpha^{(n)},~~
\tau_\alpha^{(n)}
\sim \Phi^n.\label{eq:2.10}
\end{equation}
The first term yields (\ref{eq:2.9})
\begin{equation}
\tau_\alpha^{(0)} = \Theta_\alpha\label{eq:2.11}
\end{equation}
The remaining terms are found by inserting (\ref{eq:2.10}) in (\ref{eq:2.8}),
using the relation
(\ref{eq:2.2}) and
finally identifying contributions with identical powers of $\Phi^{\alpha}$.
This
leads to the following recursion relations,
\begin{eqnarray}
\{\tau_\alpha^{(0)}(x),\tau_\beta^{(0)}(y)\}_{(\phi,\pi)} +
\{\tau_\alpha^{(1)}(x),
\tau_\beta^{(1)}(y)\}_{(\Phi)} &= 0\\\label{eq:2.12}
\{\tau_{[\alpha}^{(1)}(x),\tau_{\beta ]}^{(n+1)}(y)\}_{(\Phi)} +
B_{\alpha\beta}^{(n)}
(x,y) &=0,~~n\geq 1\label{eq:2.13}
\end{eqnarray}
with,
\begin{equation}
B_{\alpha\beta}^{(1)}(x,y)\equiv
\{\tau_{[\alpha}^{(0)}(x),\tau_{\beta]}^{(1)}(y)\}_{(\phi,
\pi)}\label{eq:2.14}
\end{equation}
\begin{eqnarray}
B_{\alpha\beta}^{(n)}(x,y) \equiv \frac{1}{2}B_{[\alpha\beta]}^{(n)}(x,y)
&\equiv&
\sum_{m=0}^n
\{\tau_\alpha^{(n-m)}(x),\tau_\beta^{(m)}(y)\}_{(\phi,\pi)}\nonumber\\
&+&\sum_{m=0}^{n-2}\{\tau_\alpha^{(n-m)}(x),\tau_\beta^{(m+2)}(y)\}_{(\Phi)},~(n\geq2)
\label{eq:2.15}
\end{eqnarray}
where the suffix $(\phi,\pi)$ or $(\Phi)$ denotes the variables with respect to
which the corresponding P.B. are evaluated. It is now straightforward to
explicitly
construct the individual terms in the series (\ref{eq:2.10}). For example,
expanding
$\tau_\alpha^{(1)}$
as,
\begin{equation}
\tau_\alpha^{(1)} (x)=\int dy X_{\alpha\beta}(x,y)\Phi^\beta(y)\label{eq:2.16}
\end{equation}
and  exploiting  (\ref{eq:2.2}),  (\ref{eq:2.11})  and (\ref{eq:2.13}) it is
found that $X_{\alpha\beta}(x,y)$ is
given by,
\begin{equation}
\int   X_{\alpha\mu}(x,z)\omega^{\mu\nu}(z,z')   X_{\beta\nu}(y,z')dz   dz'
=
-\Delta_{\alpha\beta}(x,y)\label{eq:2.17}
\end{equation}
where  $\Delta_{\alpha\beta}$  is  defined  in  (\ref{eq:2.2}).  This
determines
$\tau_\alpha^{(1)}$ (\ref{eq:2.16}).

In order to obtained the  complete  series  (\ref{eq:2.10})it  is  essential
to
introduce  the  matrix  $\omega_{\alpha\beta}$  and  $X^{\alpha\beta}$ which
are
inverse to $\omega^{\alpha\beta}$ and $X_{\alpha\beta}$ respectively,
\begin{eqnarray}
\int       \omega^{\alpha\beta}(x,y)\omega_{\beta\gamma}(y,z)        dy  &=&
\delta^\alpha_\gamma \delta(x-z)\nonumber\\
\int       X^{\alpha\beta}(x,y)X_{\beta\gamma}(y,z)        dy  &=&
\delta^\alpha_\gamma \delta(x-z)\label{eq:2.18}
\end{eqnarray}
Then  the  particular  solution of the inhomogeneous equation (\ref{eq:2.13})
is
given by,
\begin{equation}
\tau_\alpha^{(n+1)}(x) = -\frac{1}{n+2}\int \Phi^\mu(z)\omega_{\mu\nu}(z,z')
X^{
\nu\rho}(z',z'')B^{(n)}_{\rho\alpha}(z'',x)dz dz' dz''~~(n\geq1)\label{eq:2.19}
\end{equation}
The general solution to (\ref{eq:2.13}) can be obtained from  (\ref{eq:2.19})
by
adding to it a  term  containing  the  solution  of  the  homogeneous  equation
corresponding   to  (\ref{eq:2.13}).  It  has  been  shown  \cite{r8}  that
any
arbitrariness in these solutions corresponds to canonical transformations in
the
extended  phase  space. For actual computational purpose, therefore, it
suffices
to work with the solution (\ref{eq:2.16}) and (\ref{eq:2.19}).

It may be remarked that this arbitrariness in constructing the solution  of
the
first   class  constraints  is  reminiscent  of  an  analogous  feature  in
the
conventional `splitting' approach of regarding a second class system as a
gauged
fixed  first class one. The latter, it may be recalled, consists in
interpreting
one half of the second class constraints (\ref{eq:2.1}) as first class ones
and
the other half as the corresponding gauge fixing conditions. Clearly there is
an
arbitrariness in  this  splitting  which  is  also  manifested  in  the
ensuing
symplectic structure.

The  construction  of the strongly involutive set of constraints
(\ref{eq:2.10})
completes the first part of the program. It is now necessary  to  construct
the
corresponding  involutive  Hamiltonian  $\tilde  H$ which will be defined in
the
extended phase space,
\begin{equation}
\tilde H = \tilde H (\phi,\pi, \Phi) \label{eq:2.20}
\end{equation}
In the spirit of  BT's  approach,  strong  involution  (or  `abelianisation)
is
required,
\begin{equation}
\left\{\tau_\alpha (x)~,~\tilde H\right\} = 0\label{eq:2.21}
\end{equation}
subject to the boundary condition,
\begin{equation}
\tilde H (\phi, \pi, 0) = H_0 (\phi, \pi)\label{eq:2.22}
\end{equation}
where  $H_0(\phi,  \pi)$  is  the  original  Hamiltonian,  and  $\tau_\alpha$
in
(\ref{eq:2.21}) are the involutive constraints. There is a degree of freedom
in
choosing the original Hamiltonian $H_0$. This may just be the usual canonical
Hamiltonian.
Alternatively,  it may be the total Hamiltonian obtained from the canonical
part
by adding terms proportional to the second class constraints with the Lagrange
multipliers ({\it  i.e.}  the
proportionality  constants)  determined  by the usual manner of Dirac
\cite{r1}.
The particular choice of $H_0$  is crucial for technical  reasons  resulting
in
considerable  algebraic simplifications, and depends on the specific model
being
studied.

Expressing the solution to (\ref{eq:2.21}) as a power series,
\begin{equation}
\tilde  H  =  \sum_{n=0}^\infty  \tilde  H^{(n)},~~  \tilde  H^{(n)}  \sim
\Phi^n
\label{eq:2.23}
\end{equation}
with,
\begin{equation}
\tilde H^{(0)} (\phi, \pi; \Phi) = \tilde H (\phi, \pi; 0) = H_0(\phi, \pi)
\label{eq:2.24}
\end{equation}
Substituting   (\ref{eq:2.23})   and  (\ref{eq:2.10})  in  (\ref{eq:2.21}),
the
following set of recursion relations are obtained \cite{r8},
\begin{equation}
\{ \tau_\alpha^{(1)}(x) , \tilde H^{(n+1)}\}_{(\Phi)} + G_\alpha^{(n)} (x)
=0,~~
n\geq 0 \label{eq:2.25}
\end{equation}
where,
\begin{equation}
G_\alpha^{(0)}\equiv \{\tau_\alpha^{(0)}, \tilde H^{(0)}\} \nonumber
\end{equation}
\begin{equation}
G_\alpha^{(1)} \equiv \{\tau_\alpha^{(1)}, \tilde H^{(0)}\} +
 \{\tau_\alpha^{(0)}, \tilde H^{(1)}\} +
 \{\tau_\alpha^{(2)}, \tilde H^{(1)}\}_{(\Phi)}\nonumber
 \end{equation}
\begin{equation}
G_\alpha^{(n)} \equiv \sum_{m=0}^n\{\tau_\alpha^{(n-m)}, \tilde H^{(m)}\} +
 \sum_{m=0}^{n-2}\{\tau_\alpha^{(n-m)}, \tilde H^{(m+2)}\}_{(\Phi)} +
 \{\tau_\alpha^{(n+1)}, \tilde H^{(1)}\}_{(\Phi)},~~n\geq 2 \label{eq:2.26}
 \end{equation}
 It is  now  possible  to  give  the  solution  to  the  inhomogeneous
equation
(\ref{eq:2.25}),
\begin{equation}
\tilde  H^{(n+1)}  = -\frac{1}{n+1} \int \Phi^\mu(x) \omega_{\mu\nu}(x,y)
X^{\nu
\rho}(y,z) G_\rho^{(n)} (z) dx dy dz,~~(n\geq 0) \label{eq:2.27}
\end{equation}
which  gives  the  involutive  Hamiltonian.  This  completes  the  BT
\cite{r8}
construction of the first class system which is strongly involutive.

We  now  analyse  the characteristics of this construction and the need for
some
modifications. Note that the strongly involutive algebra characterises a  rank
zero
theory  \cite{r4}.  This  is  rather  restrictive since, in certain cases, a
more
natural representation could be done in terms of higher rank theories.
Although
the  rank  of  a  theory  may not be unique, yet there may be some conceptual
or
technical problems in - say - expressing a rank one theory as a rank zero
type.
Let  us illuminate by an example. It is usual to regard the Yang-Mills theory
as
of rank one since the algebra of  the  first class  (Gauss)  constraint  is
only  weakly
involutive,  expressing the standard closure property. Now it may be possible
to
write down a linear combination of the first class constraints (which would
also be
first class constraint)  so  that
these  are  strongly  involutive thereby rendering the Yang-Mills theory as
rank
zero. In that case, however, apart from arcane algebraic structures  the
natural
identity  of the first class constraint with the zero component of the equation
of motion gets lost
and its simple interpretation as the generator of the gauge  transformation
may
also  be  difficult  to preserve. Keeping these points in view, therefore, it
is
desirable to extend the BT  formalism  to  represent  higher  rank  theories
by
modifying the involutive algebra (\ref{eq:2.8}) as ,
\begin{equation}
\{\tau_\alpha^a(x),\tau_\beta^b(y)\}=f^{abc}\int
\Gamma_{\alpha\beta\gamma}(x,y,z) \tau_\gamma^c(z) dz\label{eq:2.28}
\end{equation}
where the structure functions are field independent. This characterises  a
rank
one  theory. A rank two theory would occur if the structure functions are field
dependent. This happens in the case of gravity which is kept outside  the
realm
of  the  present  paper.  The  algebra  (\ref{eq:2.28})  is ideal for
discussing
nonabelian theories where  $a$,  $b$,  $c$  represent  the  corresponding
group
indices.   Note   that,  in  the  abelian  case,  (\ref{eq:2.28})  goes  over
to
(\ref{eq:2.8}).

Following similar techniques as BT \cite{r8} by expressing $\tau_\alpha^a(x)$
as
a power series analogous to (\ref{eq:2.10}),
\begin{equation}
\tau_\alpha^a  =  \sum_{n=0}^\infty \tau_\alpha^{a(n)};
{}~~\tau_\alpha^{a(n)}\sim
(\Phi^{\alpha a})^n \label{eq:2.29}
\end{equation}
we obtain the following relations corresponding to
(\ref{eq:2.12})--(\ref{eq:2.15})
with the appropriate group indices,
\begin{eqnarray}
\{ \tau_\alpha^{a(0)} (x), \tau_\beta^{b(0)}(y) \} &+&
\{ \tau_\alpha^{a(1)} (x), \tau_\beta^{b(1)}(y) \}_{(\Phi)}\nonumber\\
&=& f^{abc} \int \Gamma_{\alpha\beta\gamma}(x,y,z) \tau_\gamma^{c(0)}(z) dz
\label{eq:2.30a}
\end{eqnarray}
\begin{equation}
\{\tau_{[\alpha}^{a(1)}(x),        \tau_{\beta]}^{b(n+1)}(y)\}_{(\Phi)}
+
B_{\alpha\beta}^{ab(n)}(x,y) = 0,~~n\geq 1 \label{eq:2.30b}
\end{equation}
where,
\begin{equation}
B_{\alpha\beta}^{ab(1)}(x,y)=
\{ \tau_{[\alpha}^{a(0)} (x), \tau_{\beta]}^{b(1)}(y) \}
- f^{abc} \int \Gamma_{\alpha\beta\gamma}(x,y,z) \tau_\gamma^{c(1)}(z) dz
\label{eq:2.30c}
\end{equation}
\begin{eqnarray}
B_{\alpha\beta}^{ab(n)}(x,y)&=&
\sum_{m=0}^n\{ \tau_\alpha^{a(n-m)} (x), \tau_\beta^{b(m)}(y) \} +
\sum_{m=0}^{n-2}\{ \tau_\alpha^{a(n-m)} (x), \tau_\beta^{b(m+2)}(y)
\}_{(\Phi)}\nonumber\\
&-&  f^{abc}  \int \Gamma_{\alpha\beta\gamma}(x,y,z) \tau_\gamma^{c(n)}(z)
dz,~~
n\geq2
\label{eq:2.30d}
\end{eqnarray}
It may be checked that the  required  solution  of  the  inhomogeneous
equation
(\ref{eq:2.30b}) is given by,
\begin{equation}
\tau_\alpha^{a(0)} = \Theta_\alpha^a \nonumber
\end{equation}
\begin{equation}
\tau_\alpha^{a(1)} (x)=\int dy
X_{\alpha\beta}^{ab}(x,y)\Phi^{b\beta}(y)\label{eq:2.31}
\end{equation}
with,
\begin{eqnarray}
\int   X_{\alpha\mu}^{ac}(x,z)\omega^{\mu\nu}_{cd}(z,z')   &
&X_{\beta\nu}^{bd}(y,z')dz  dz'\nonumber\\
&-& f^{abc}\int \Gamma_{\alpha\beta\gamma}(x,y,z) \Theta_\gamma^c(z)=
-\Delta_{\alpha\beta}^{ab}(x,y)\label{eq2.32}
\end{eqnarray}
and,
\begin{eqnarray}
\tau_\alpha^{a(n+1)}(x) = -\frac{1}{n+2}\int \Phi^{\mu
b}(z)\omega_{\mu\nu}^{bc}(z,z') X^{
\nu\rho}_{cd}(z'&,&z'')B^{da(n)}_{\rho\alpha}(z'',x)dz dz' dz'' \nonumber\\
& &(n\geq1)\label{eq:2.33}
\end{eqnarray}
where $\Delta_{\alpha\beta}^{ab}$, $\omega_{\mu\nu}^{ab}$, $X_{ab}^{\mu\nu}$
are
the  analogues  of  (\ref{eq:2.2}),  (\ref{eq:2.18})  with the appropriate
group
indices.  This  completes  the  constructions  of  the  involutive
constraints
(\ref{eq:2.29})  satisfying  the algebra (\ref{eq:2.28}). A similar analysis
can
be done for the Hamiltonian which is straightforward and not repeted.
\vspace{1cm}
\section {\bf An abelian example-- the Proca model}

It is instructive to apply the general ideas discussed in the  previous
section
to specific models. Such model based analysis clarifies several issues and
gives
a deeper insight into the general formalism. Let  us  first  consider  a
simple
abelian  example  which  is  the  Proca  model whose dynamics is governed by
the
Lagrangian density,
\begin{equation}
{\cal L}  =  -\frac{1}{4}  F_{\mu\nu}F^{\mu\nu}  +  \frac{1}{2}  m^2  A^\mu
A_\mu
\label{eq:3.1}
\end{equation}
Note the introduction of an explicit mass term which breaks the gauge
invariance
of the usual Maewell's theory. Consequently  (\ref{eq:3.1}) represents a
second
class  system  which  is  easily  confirmed by the standard constraint
analysis.
There is one primary constraint,
\begin{equation}
\Theta_1 = \pi_0\approx 0 \label{eq:3.2a}
\end{equation}
and one secondary constraint,
\begin{equation}
\Theta_2 = \partial_i\pi^i + m^2 A_0\approx 0 \label{eq:3.2b}
\end{equation}
obtained by conserving $\Theta_1$ with the total Hamiltonian,
\begin{equation}
H_T = H_c + \int d^3x \lambda \Theta_1 \label{eq:3.3}
\end{equation}
where $H_c$ is the canonical Hamiltonian,
\begin{equation}
H_c = \int \left[\frac{1}{2} \pi_i^2 + \frac{1}{4} F_{ij}^2 + \frac{1}{2}
(A_0^2
+ A_i^2) - A_0\Theta_2 \right] \label{eq:3.4}
\end{equation}
and  $\lambda$  is  a  Lagrange  multiplier,  while  $\pi_\mu$  is  the
momenta
canonically conjugate to $A^\mu$. The algebra,
\begin{equation}
\Delta_{\alpha\beta}   (x,y)   =   \{\Theta_\alpha,   \Theta_\beta\}    =
m^2
\epsilon_{\alpha\beta} \delta(x-y); ~~\alpha, \beta = 1,2 \label{eq:3.5}
\end{equation}
where the antisymmetric tensor $\epsilon_{\alpha\beta}$ is normalised as,
\begin{equation}
\epsilon_{12} = - \epsilon^{12} = -1 \label{eq:3.6}
\end{equation}
reveals the second class nature of the constraints $\Theta_\alpha (x)$.

In  order  to  convert  this  system into first class, the first objective is
to
transform $\Theta_\alpha$  into  first  class  by  extending  the  phase
space.
Following  the  general approach we have to introduce the matrix
(\ref{eq:2.5}),
which is chosen as,
\begin{equation}
\omega^{\alpha\beta} (x,y) = 2\epsilon^{\alpha\beta}\delta(x-y) \label{eq:3.7}
\end{equation}
Then the other matrix $X_{\alpha\beta}$ in the   game  is  obtained  by
solving
(\ref{eq:2.17}) with $\Delta_{\alpha\beta}$ given by (\ref{eq:3.5}),
\begin{equation}
X_{\alpha\beta} (x,y) = \left(
\begin{array}{lr}
1&0\nonumber\\ 0& \frac{m^2}{2}\nonumber
\end{array}
\right) \delta(x-y) \label{eq:3.8}
\end{equation}
There  is  an  arbitrariness  in  choosing $\omega^{\alpha\beta}$
(\ref{eq:3.7})
which would naturally be manifested in (\ref{eq:3.8}). This just corresponds
to
canonical transformations in the extended phase space. However, as has also
been
evidenced in other calculations \cite{r18,r19,r20}, the choice (\ref{eq:3.7})
brings about
remarkable algebraic simplifications.

Using  (\ref{eq:2.10}),  (\ref{eq:2.16})  and  (\ref{eq:3.8})  the  new  set
of
constraints are found to be,
\begin{eqnarray}
\tau_1 &=& \Theta_1 + \Phi^1\nonumber\\
\tau_2 &=& \Theta_2 + \frac{m^2}{2} \Phi^2 \label{3.9}
\end{eqnarray}
which are strongly involutive,
\begin{equation}
\{\tau_\alpha, \tau_\beta\} = 0 \label{3.10}
\end{equation}
Recall the $\Phi^1$, $\Phi^2$ are  the  new  variables  satisfying  the
algebra
(\ref{eq:2.4})  with  $\omega^{\alpha\beta}$  given  by (\ref{eq:3.7}). Observe
further that only $\tau_\alpha^{(1)}$ (\ref{eq:2.16}) contributes in the series
(\ref{eq:2.10})  defining  the  first class constraint.   All   higher   order
 terms   given   by
(\ref{eq:2.19})   vanish   as   a   consequence   of   our   intelligent
choice
(\ref{eq:3.7}).

The next step is to obtain the involutive Hamiltonian. The generating
functional
$G_\alpha^{(n)}$   which   determines   this   Hamiltonian   is   obtained
from
(\ref{eq:2.26}),
\begin{equation}
G_\alpha^{(0)} = \{\Theta_\alpha,H_c\} \nonumber
\end{equation}
\begin{equation}
G_\alpha^{(n)} = \{\tau_\alpha^{(1)},\tilde H^{(n-1)}\} +
\{\Theta_\alpha,\tilde H^{(n)}\},~~(n\geq 1) \label{3.11}
\end{equation}
where $\tilde H^{(n)}$ is given in (\ref{eq:2.27}) and the original
Hamiltonian
$\tilde  H^{(0)}$  is taken to be the canonical Hamiltonian (\ref{eq:3.4}). It
is
noteworthy that the general expression (\ref{eq:2.26}) reduces to the
remarkably
simple form (\ref{3.11}) since there are {\it only} two terms $\Theta_\alpha$
and
$\tau_\alpha^{(1)}$ in the expansion (\ref{eq:2.10}), which is the result of
the
judicious choice (\ref{eq:3.7}). Explicit calculations yield,
\begin{equation}
G_1^{(0)} = \Theta_2 \nonumber
\end{equation}
\begin{equation}
G_2^{(0)} = m^2 \partial_iA^i. \label{3.12}
\end{equation}
which is substituted in (\ref{eq:2.27}) to obtain $\tilde H^{(1)}$,
\begin{equation}
\tilde    H^{(1)}   =   \int   \left[   (\partial_iA^i)   \Phi^1   -
\frac{1}{2}
(\partial_i\pi^i + m^2 A_0) \Phi^2\right] d^3x \label{3.13}
\end{equation}
This is inserted back in (\ref{3.11})  to  deduce  $G_\alpha^{(1)}$  which
then
yields $\tilde H^{(2)}$ from (\ref{eq:2.27}),
\begin{equation}
\tilde  H^{(2)}  =  \int  d^3x \left[ -\frac{m^2}{8} (\Phi^2)^2 -
\frac{1}{2m^2}
(\partial_i\Phi^1)(\partial^i\Phi^1)\right] \label{3.14}
\end{equation}
after which the series terminates. Thus the complete expression for the
desired
Hamiltonian is,
\begin{equation}
\tilde H = H_c + \tilde H^{(1)} + \tilde H^{(2)} \label{3.15}
\end{equation}
which, by construction, is strongly involutive,
\begin{equation}
\{\tilde H, \tau_\alpha\} = 0 \label{3.16}
\end{equation}
This  completes  the  operatorial  (`abelian') conversion of the original
second
class system (with Hamiltonian $H_c$ and constraints $\Theta_\alpha$) into
first
class (with Hamiltonian $\tilde H$ and constraints $\tau_\alpha$).

It  is  well  known,  however,  that  there  exists  the  St\"uckelberg
\cite{r9}
mechanism whereby the second class Lagrangian (\ref{eq:3.1}) can be embedded
into
the corresponding first class theory by extending the configuration space,
\begin{equation}
{\cal   L}^\prime   =   -\frac{1}{4}  F_{\mu\nu}^2  +  \frac{1}{2}m^2  (A_\mu
+
\partial_\mu\theta)(A^\mu + \partial^\mu\theta) \label{3.17}
\end{equation}
where $\theta$, the new field,  is  the  St\"uckelberg  scalar.  The
Lagrangian
(\ref{3.17}) is invariant under the gauge transformations,
\begin{eqnarray}
A_\mu &\to& A_\mu - \partial_\mu\alpha\nonumber\\
\theta &\to& \theta + \alpha \label{3.18}
\end{eqnarray}
and characterises a first class theory.

We   now   unravel   the   correspondence   of  the  Hamiltonian  approach
with
St\"uckelberg's formalism. The first step  is  to  identify  the  new
variables
$\Phi^1$,  $\Phi^2$  (in  the  Hamiltonian formalism) as a canonically
conjugate
pair ($\rho$, $\pi_\rho$),
\begin{eqnarray}
\Phi^1 &\to& m^2 \rho\nonumber\\
\Phi^2 &\to& \frac{2}{m^2}\pi_\rho  \label{3.19}
\end{eqnarray}
as may be easily checked from (\ref{eq:2.4}) and (\ref{eq:3.7}). The phase
space
partition function is then given by the Faddeev formula \cite{r4},
\begin{equation}
Z=  \int  {\cal  D}(A_\mu  \pi^\mu  \rho  \pi_\rho) \prod_{\alpha,\beta}
\delta(
\tau_\alpha)   \delta(\Gamma_\beta)   \det\{
\tau_\alpha,\Gamma_\beta\}e^{iS}
\label{3.20a}
\end{equation}
where,
\begin{equation}
S  =  \int \left( \pi_\mu {\dot A}^\mu + \pi_\rho {\dot \rho} - \tilde H
\right)
\label{3.20b}
\end{equation}
with $\tilde H$, the involutive Hamiltonian (\ref{3.15}), now expressed in
terms
of  $(\rho,  \  \pi_\rho)$  instead  of $\Phi^1$, $\Phi^2$. The gauge
conditions
$\Gamma_\beta$ are chosen so that the determinant occurring  in  the
functional
measure  is  nonvanishing. Moreover $\Gamma_\beta$ are assumed to be
independent
of the momenta so that these may  be  considered  as  Faddeev-Popov  type
gauge
conditions.  We  now  perform  the  momentum  integrations  to  pass  on  to
the
configuration space  partition  function.  The  $\pi_0$  integral  is
trivially
performed  by exploiting the delta function $\delta(\tau_1) = \delta(\pi_0 +
m^2
\rho)$ in  (\ref{3.20a}).      The  other  delta  function  $\delta(\tau_2)$
is
expressed  by  its  Fourier  transform  (with  Fourier  variable  $\xi$) and
the
Gaussian integral over $\pi_\rho$ performed. This yields the action,
\begin{eqnarray}
S &=&  \int  [\pi_i  {\dot  A}^i - \xi \partial_i\pi^i + \frac{m^2}{2} ({\dot
\rho}^2
+\xi^2 - 2{\dot \rho}\xi) - \frac{1}{2}\pi_i^2 + \frac{1}{4}  F_{ij}^2
\nonumber
\\     &+&   \frac{1}{2}m^2   A_i^2   +m^2   \partial_iA^i\rho   -
\frac{m^2}{2}
\partial_i\rho \partial^i\rho ] \label{3.21}
\end{eqnarray}
Note that the $A_0$ term cancelled out.  The  integral  over  $A_0$,
therefore,
reduces  to  a trivial identity since one of the gauge conditions
$\Gamma_\beta$
must involve $A_0$ to have a nonvanishing P.B. with the constraint $\tau_1$.
The
Gaussian  integral  over  $\pi_i$  is finally performed and the Fourier
variable
$\xi$ relabelled as ($-A_0$) to express the action in a covariant form,
$$
S=\int[- \frac{1}{4}  F_{\mu\nu}^2
+  \frac{1}{2}m^2   A_\mu^2   -m^2 \partial_\mu A^\mu\rho  + \frac{m^2}{2}
\partial_\mu\rho \partial^\mu\rho ] \nonumber
$$
\begin{equation}
=\int[- \frac{1}{4}  F_{\mu\nu}^2
+  \frac{1}{2}m^2   (A_\mu+\partial_\mu\rho)^2   -m^2 \partial_\mu(A^\mu\rho)]
\label{3.22}
\end{equation}
Ignoring the last (boundary) term  we find that the Lagrangian corresponding
to
(\ref{3.22})   agrees   with   the  usual  St\"uckelberg  form  (\ref{3.17})
by
identifying $\rho$ with the St\"uckelberg field $\theta$.  There  are,
however,
some  interesting  points concerning this correspondence between the
Hamiltonian
and Lagrangian embeddings which are now elaborated.

The  first  observation  is  that  St\"uckelberg  Lagrangian  (\ref{3.17})
(or,
equivalently,  considering  (\ref{3.22}) without the boundary term) leads to
the
primary constraint,
\begin{equation}
T_1'= \pi_0\approx 0 \label{3.23}
\end{equation}
The canonical Hamiltonian obtained by a Legendre transform is,
\begin{equation}
H_c' = \int  [  \frac{1}{2}  \pi_i^2  +  \frac{\pi_\rho^2}{2m^2}  +
\frac{1}{4}
F_{ij}^2 + \frac{m^2}{2}(A_i + \partial_i\rho)^2 - A_0 T_2'] \label{3.24}
\end{equation}
where $T_2'$ is the secondary constraint,
\begin{equation}
T_2' = \partial_i\pi^i + \pi_\rho \approx 0 \label{3.25}
\end{equation}
No further constraints are present. It simple to see that $T_1'$ and $T_2'$
form
a pair of first class constraints which are in (weak) involution with the
total  Hamiltonian.  The
system is first class which is precisley what one expects. The point to be
emphasised is,
however, that  the  set  of  constraints  $(T_1',  T_2')$  and  the
Hamiltonian
(\ref{3.24})  do  {\it not} agree with the corresponding structures (\ref{3.9},
\ref{3.15}, \ref{3.19}) obtained in the Hamiltonian approach. In the  latter
it  is  necessary  to  modify  the  structure  of {\it all} the original
second
class constraints in
(\ref{3.9}) to make them first class. Here,  however,  only  the  `Gauss'
constraint  is
modified  (\ref{3.25})  while  the  other  (\ref{3.23}) is not. Indeed it can
be
checked  that  in  the  approach  of  BT  \cite{r8}  a  partial  modification
of
(cf. (\ref{3.23}),   (\ref{3.25}))  would  lead  to  an  algebraic
inconsistency.  More
specifically,   the   matrix   $X_{\alpha\beta}$   (\ref{eq:2.17})   cannot
be
constructed. Consequently it is found that although it is possible to obtain
the
St\"uckelberg form (\ref{3.17}) starting from the BT construction, the
converse
is not true.

The  next  question, naturally, is to find the Lagrangian compatible with the
BT
construction.  Indeed  we  show  that  it  is  just  the  first  expression
in
(\ref{3.22}),
 \begin{equation}
{\cal L}''=- \frac{1}{4}  F_{\mu\nu}^2
  +\frac{1}{2}m^2         A_\mu^2         -m^2       \partial_\mu
A^\mu\rho
+\frac{m^2}{2}(\partial_\mu\rho)^2
\label{3.26}
\end{equation}
The primary constraint is,
\begin{equation}
T_1'' = \pi_0 + m^2 \rho \approx 0 \label{3.27}
\end{equation}
The canonical Hamiltonian obtained from (\ref{3.26}) is
\begin{eqnarray}
H &=& + \frac{1}{2}\pi_i^2+\frac{\pi_\rho^2}{2m^2}
+ \frac{1}{4}  F_{ij}^2 - \frac{1}{2}m^2   A_\mu^2\nonumber\\
&+& m^2   \partial_iA^i\rho   -   \frac{m^2}{2}
\partial_i\rho \partial^i\rho -A_0\partial_i\pi^i \label{3.28}
\end{eqnarray}
Time conserving $T_1''$ (\ref{3.27}) leads to the secondary constraint,
\begin{equation}
T_2'' = \partial_i\pi^i + m^2 A_0 + \pi_\rho \approx 0 \label{3.29}
\end{equation}
It  can  be  checked that no further constraints are generated by this
iterative
scheme. Furthermore $T_1''$, $T_2''$ are a pair of first class constraints  in
involution  with  the
Hamiltonian  (\ref{3.28}).  It  is important to note further that $T_1''$,
$T_2''$
are exactly identical to the set of first class constraints (\ref{3.9},
\ref{3.19}) obtained in the
Hamiltonian formalism. Moreover the first class Hamiltonian (\ref{3.28})
differs from the
involutive  Hamiltonian  (\ref{3.15}) by  a term proportional to the first
class
constraint  $\tau_2$
(\ref{3.9}),
\begin{equation}
H = \tilde H + \frac{\pi_\rho}{m^2} \tau_2 \label{3.30}
\end{equation}
Acting  on  physical  states,  this  difference is trivial since such states
are
annihilated by the first class constraints. Similarly the equations of motion
for observable  ({\it
i.e.}  gauge  invariant  variables)  will  also be unaffected by this
difference
since $\tau_2$ can be regarded as the generator of  the  gauge
transformations.
In   the   construction   of   the   functional   integral  this  difference
is
inconsequential since the constraint $\tau_2$ is  strongly  implemented  by
the
delta  function  $\delta(\tau_2)$  (\ref{3.20a}). Thus $H$ and $\tilde H$ may
be
regarded as canonically equivalent. This  completes  our  demonstration  of
the
compatibility of the Lagrangian (\ref{3.26}) with the Hamiltonian description
of
BT.

To  summarise, we emphasise the role played by the apparantly innocuous
boundary
term in (\ref{3.22}). If  we  drop  it,  the  resulting  Lagrangian  is
exactly
identical  to  the  St\"uckelberg form (\ref{3.17}) with the BT field
identified
with the St\"uckelberg  scalar. In that case, however, the Hamiltonian  and
the
set  of constraints obtained from this Lagrangian are completely inequivalent
to
the original BT construction. If, on the contrary, the boundary term is
retained
it  yields the embedded Lagrangian (\ref{3.26}). The constraints and
Hamiltonian
following from  this  Lagrangian  are  completely  equivalent  to  the
original
Hamiltonian  embedding.  In this case, therefore, the cycle - Hamiltonian to
the
Lagrangian to the Hamiltonian - closes.

\section {\bf A nonabelian example -- The nonabelian Proca model}

In  this  section  we  consider  the  implication  of  the generalised
canonical
approach for a nonabelian model whose dynamics is  governed  by  the
Lagrangian
density,
\begin{equation}
{\cal  L}  =  -  \frac{1}{4}  F^a_{\mu\nu} F^{\mu\nu,a} + \frac{1}{2}m^2
A_\mu^a
A^{\mu,a} \label{4.1}
\end{equation}
which is just the nonabelian extension of the usual Proca model
(\ref{eq:3.1}).
The corresponding gauge group may be arbitrarily taken whose structure
constants
will be denoted by $f^{abc}$. There is a primary constraint,
\begin{equation}
T_1^a = \pi_0^a \approx 0\label{4.2}
\end{equation}
which, conserved with the total Hamiltonian,
\begin{equation}
H_T = H_c + \int u^a(x)\pi_0^a(x) \label{4.3}
\end{equation}
where $H_c$ is the canonical Hamiltonian,
\begin{equation}
H_c  =  \int  [ \frac{1}{2} (\pi_i^a)^2 + \frac{1}{2}m^2(A_i^a)^2 + \frac{1}{4}
(F_{ij}^a)^2 + \frac{m^2}{2}(A_0^a)^2 - A_0^a T_2^a ] \label{4.4}
\end{equation}
yields a secondary constraint,
\begin{equation}
T_2^a = \partial_i\pi^{ia} - g f^{abc}\pi^{ib}A_i^c +m^2 A_0^a \label{4.5}
\end{equation}
The Poisson algebra of constraints,
\begin{equation}
\{T_1^a(x),T_1^b(y)\} = 0\label{4.6}
\end{equation}
\begin{equation}
\{T_1^a(x),T_2^b(y)\} = -m^2\delta^{ab}\delta(x-y)\label{4.7}
\end{equation}
\begin{equation}
\{T_2^a(x),T_2^b(y)\}   =   g   f^{abc}T_2^c\delta(x-y)   -   gm^2f^{abc}
A_0^c
\delta(x-y)\label{4.8}
\end{equation}
clearly   illustrates   that  $T_1^a$,  $T_2^a$  are  a  set  of  second  class
constraints. Note, particularly the equation (\ref{4.8}). This algebra has been
given
incorrectly by Senjanovic \cite{r15} where the second term in  the  right  hand
side  of  (\ref{4.8})  is missing. Cosequently the corresponding Dirac Brackets
(involving $A_0^a$) have been {\it incorrectly} evaluated. The correct Dirac
brackets may
be computed from (\ref{4.6}-\ref{4.8}) and are found to be,
\begin{equation}
\{\pi^{ia}(x),  A_0^b(y)\}_{DB}  =  -\frac{g}{m^2}  f^{abc}\pi^{ic}(x)
\delta(x-y)
\nonumber
\end{equation}
\begin{equation}
\{\pi^{ia}(x),  A_j^b(y)\}_{DB} = -\delta^{ab}\delta^i_j \delta(x-y)\nonumber
\end{equation}
\begin{equation}
\{\pi^{ia}(x),  \pi_j^b(y)\}_{DB} = 0\nonumber
\end{equation}
\begin{equation}
\{A^{ia}(x),  A_j^b(y)\}_{DB} = 0\nonumber
\end{equation}
\begin{equation}
\{A^a_0(x),  A_j^b(y)\}_{DB} = \frac{1}{m^2}\partial_j^x\delta^{ab}  \delta
(x-y)
-\frac{g}{m^2}f^{abc} A_j^c(x)\delta (x-y)\nonumber
\end{equation}
\begin{equation}
\{A_0^a(x)   ,   A_0^b(y)\}_{DB}  =  -\frac{g}{m^2}f^{abc}  A_0^c(x)\delta
(x-y).
\label{4.n}
\end{equation}

There is one important difference in the Dirac brackets of the nonabelian
theory
(\ref{4.1}) contrasted with the corresponding abelian  version  (\ref{eq:3.1}).
In  the  former  case  these  brackets are {\it field dependent}. Consequently
a
transition to quantum theory where these DBs are replaced by commutators will
be
problematic \cite{r1}. this was also
mentioned  earlier in our introduction. In this case, therefore, the
generalised
canonical formalism provides a viable alternative to quantisation. By
converting
the  theory  into  first  class,  the  need for Dirac brackets is eliminated
and
quantisation can proceed by using the canonical Poisson brackets.

As analysed in
the  previous section, the initial step is to obtain the first class
constraints
from the second class ones, (\ref{4.2}) and (\ref{4.5}).
In order to construct the first class constraints $\tau_\alpha^a$ satisfying
the
involutive algebra (\ref{eq:2.28}) we have to  specify  the  matrices
$\omega^{
\alpha\beta}_{ab}$, $X_{\alpha\beta}^{ab}$ which are the nonabelian analogues
of
(\ref{eq:3.7}) and (\ref{eq:3.8})respectively. We make the following choice,
\begin{equation}
\omega^{\alpha\beta}_{ab}(x,y)=
m^2\epsilon^{\alpha\beta}\delta^{ab}\delta(x-y)
\label{4.10}
\end{equation}
\begin{equation}
X_{\alpha\beta}^{ab}(x,y) = \left(
\begin{array}{lr}
2\delta^{ab} & 0\nonumber \\ gf^{abc}A_0^c & \frac{1}{2}\delta^{ab}\nonumber
\end{array} \right)\delta(x-y)\label{4.11}
\end{equation}
The corresponding inverse matrices are,
\begin{equation}
\omega_{\alpha\beta}^{ab}(x,y)=\frac{1}{m^2}\epsilon_{\alpha\beta}\delta_{ab}\delta(x-y)
\label{4.12}
\end{equation}
\begin{equation}
X^{\alpha\beta}_{ab}(x,y) = \left(
\begin{array}{lr}
\frac{1}{2}\delta^{ab} & 0\nonumber \\ -gf^{abc}A_0^c & 2\delta^{ab}\nonumber
\end{array} \right)\delta(x-y)\label{4.13}
\end{equation}
We  further  take  the  following  form  for
$\Gamma_{\alpha\beta\gamma}(x,y,z)$
appearing in (\ref{eq:2.28}),
\begin{equation}
\Gamma_{\alpha\beta\gamma}(x,y,z) = g \delta_{\alpha 2}\delta_{\beta 2}
\delta_{
\gamma 2}\delta(x-y)\delta(y-z)\label{4.14}
\end{equation}
so  that  (\ref{eq:2.28})  with  (\ref{4.14}) yields the conventional
involutive
algebra valid for the first class Yang-Mills theory.

With the choice (\ref{4.10}), (\ref{4.11}), (\ref{4.14})  one  can  compute
the
generating functions $B^{ab}_{\alpha\beta}$ (\ref{eq:2.30a}-\ref{eq:2.30d}).
{}From this knowledge
the various terms (\ref{eq:2.33}) in the power series  expansion
(\ref{eq:2.29})
may  be  obtained.  We  give  below  the  final  expressions for the first
class
constraints  $\tau^a_1$,  $\tau_2^a$  satisfying  (\ref{eq:2.28}),
(\ref{4.14})
valid upto the second power in the additional fields $\Phi^{\alpha ,a}$:
\begin{equation}
\tau_1^a = T_1^a + 2 \Phi^{1,a} \label{4.15}
\end{equation}
\begin{eqnarray}
\tau_2^a  =  T_2^a  &+&  \frac{1}{2}  \Phi^{2,a}  +  g  f^{abc}
\Phi^{1,b}A_0^c -
\frac{g}{2m^2} f^{abc}\Phi^{2,b}\Phi^{1,c}\nonumber\\
&+&  \frac{g^2}{3m^2}  f^{adb}f^{bce}A_0^e\Phi^{1,c}\Phi^{1,d}  +
O(\Phi\Phi\Phi)
\label{4.16}
\end{eqnarray}
where $T_1^a$, $T_2^a$ are the original second class constraints. Note further
that
$\tau_1^a$  is  an  {\it  exact}  result  while  $\tau_2^a$ gets modified in
the
different powers of the new fields $\Phi$.

The  next step is to compute the involutive Hamiltonian. To complete the
analogy
with the usual Yang-Mills system we demand the following algebra to be
satisfied
by the involutive Hamiltonian $\tilde H$ and the constraints,
\begin{eqnarray}
\{\tau_1^a (x), \tilde H\} &=& \tau_2^a(x)\nonumber\\
\{\tau_2^a(x), \tilde H\} &=& gf^{abc}A_0^b\tau_2^c\label{4.17}
\end{eqnarray}
As  usual  $\tilde  H$ is expressed as a power series (\ref{eq:2.23}), with the
original Hamiltonian $H_0$ taken  to  be  the  canonical  piece.  For
algebraic
considerations  we  mention  that  the  old  (second  class) constraints and
the
original (canonical) Hamiltonian have identical brackets as  (\ref{4.17}),
{\it
i.e.}
\begin{eqnarray}
\{T_1^a (x),  H_c\} &=& T_2^a(x)\nonumber\\
\{T_2^a(x),  H_c\} &=& gf^{abc}A_0^b(x)T_2^c(x)\label{4.18}
\end{eqnarray}

We  now wish to compute the generating function $G_\alpha^{(n)}$
(\ref{eq:2.26})
which will yield the desired involutive Hamiltonian. Recall that  the
structure
given  in  (\ref{eq:2.26}) leads to a strongly involutive Hamiltonian
satisfying
(\ref{eq:2.21}). Since our  algebra  (\ref{4.17})  is  only  weakly
involutive,
suitable  modifications  must be made in the analysis given in section 2. Let
us
first express (\ref{4.17}) in a covariant notation,
\begin{equation}
\{ \tau_\alpha^a(x), \tilde H\}  =  \int dy
V^{ab}_{\alpha\beta}(x,y)\tau_\beta^b(y)
\label{4.19}
\end{equation}
so that,
\begin{eqnarray}
V_{12}^{ab}(x,y) &=& \delta^{ab}\delta(x-y)\nonumber\\
V_{22}^{ab} (x,y) &=& -gf^{abc}A_0^c\delta(x-y) \label{4.20}
\end{eqnarray}
with all other $V$-coefficients being zero.

Proceeding  as  was  done  in sec 2, the modified expressions for the
generating
functions (\ref{eq:2.26}) are found to be,
\begin{equation}
G_\alpha^{a(0)}\equiv \{\tau_\alpha^{a(0)}, \tilde H^{(0)}\}
-\int dy V_{\alpha\beta}^{ab}(x,y) \tau_\beta^{b(0)}(y)\nonumber
\end{equation}
\begin{eqnarray}
G_\alpha^{a(1)} \equiv \{\tau_\alpha^{a(1)}, \tilde H^{(0)}\} &+&
 \{\tau_\alpha^{a(0)}, \tilde H^{(1)}\} +
 \{\tau_\alpha^{a(2)}, \tilde H^{(1)}\}_{(\Phi)}\nonumber\\
&-&\int dy V_{\alpha\beta}^{ab}(x,y) \tau_\beta^{b(1)}(y)\nonumber
 \end{eqnarray}
\begin{eqnarray}
G_\alpha^{a(n)} \equiv \sum_{m=0}^n\{\tau_\alpha^{a(n-m)}, \tilde H^{(m)}\} &+&
 \sum_{m=0}^{n-2}\{\tau_\alpha^{a(n-m)}, \tilde H^{(m+2)}\}_{(\Phi)} +
 \{\tau_\alpha^{a(n+1)}, \tilde H^{(1)}\}_{(\Phi)}\nonumber\\
 &-&\int dy V_{\alpha\beta}^{ab}(x,y) \tau_\beta^{b(n)}(y), ~~n\geq 2
\label{4.21}
 \end{eqnarray}
The final solution for $\tilde H$ satisfying (\ref{4.19}) is given by the
power
series (\ref{eq:2.23}) with,
\begin{equation}
\tilde   H^{(n+1)}   =   -\frac{1}{n+1}   \int
\Phi^{\mu,a}\omega^{ab}_{\mu\nu}
X^{\nu\rho}_{bc}G^{c(n)}_{\rho},~~(n\geq 0) \label{4.22}
\end{equation}
where $\omega^{ab}_{\mu\nu}$, $X^{\nu\rho}_{bc}$ are  defined  in
(\ref{4.12}),
(\ref{4.13}).  The  result  for  the  involutive  Hamiltonian  now  follows by
a
straightforward algebra,
\begin{equation}
\tilde H = H_c -\frac{1}{m^2}\int \Phi^{1a}\Phi^{2a}  -  \frac{2g}{m^2}
f^{abc}
f^{cge} \int A_0^b A_0^g\Phi^{1a}\Phi^{1e} + O(\Phi\Phi\Phi) \label{4.23}
\end{equation}
correct to second powers in the new fields.

It is now possible to establish a connection of our analysis with the
Lagrangian
formalism \cite{r16} of converting (\ref{4.1}) into first class.  In
\cite{r16}
the  gauge  group  has  been chosen as $SU(2)$ for algebraic simplification.
The
configuration space is extended by introducing three additional fields which
are
just the Eulerian angles \cite{r16}. This implies the addition of six new
fields
in the phase space (the fields of the configuration space and their
canonically
conjugate  momenta).  The  new fields that we have introduced are precisely six
in
number for the group $SU(2)$. Three are  given  by  $\Phi^{1,a}$  and  three
by
$\Phi^{2,a}$ ($a$=1, 2, 3). The involutive algebra in the Lagrangian
formulation
\cite{r16}     is     exactly     identical     to     (\ref{eq:2.28})
with
$\Gamma_{\alpha\beta\gamma}$   specified   by  (\ref{4.14}),  and
(\ref{4.17}).
Conceptually, therefore, the first class system found here  by  the
Hamiltonian
formalism  corresponds  to  the  Lagrangian  scheme \cite{r16}. To make an
exact
correspondence  which  will  equate  some  combination  of   the
$\Phi^{1,a}$,
$\Phi^{2,a}$ fields with the Eulerian fields (and their conjugates) is more of
a
technical problem. We will have  to  isolate  the  terms  in  the  power
series
(\ref{4.16}),  (\ref{4.23})  which  will  yield  the  corresponding terms in
the
trigonometric (sine, tan etc) expansions given in \cite{r16}.

We  will close this section by pointing out certain erroneous observations made
in the literature \cite{r17} regarding the conversion of (\ref{4.1}) to a first
class
system by adopting the usual St\"uckelberg \cite{r9}  mechanism.  Just  as  the
abelian  (Proca)  model  (\ref{eq:3.1})  could be converted into first class by
making a `gauge transformation' (see  ),
\begin{equation}
A_\mu \to A_\mu + \partial_\mu \theta \label{4.24}
\end{equation}
it has been claimed \cite{r17} that a `nonabelian gauge transformation'
\begin{equation}
A_\mu^a \to A_\mu^a + (D_\mu\theta)^a \label{4.25}
\end{equation}
is able to transform (\ref{4.1}) into a first class system. This is wrong as we
promptly
demonstrate.

The Lagrangian (\ref{4.1}) modified by (\ref{4.25}) is,
\begin{equation}
{\cal L} = -\frac{1}{4} (F_{\mu\nu}^a)^2 + \frac{m^2}{2}(A_\mu^a+  (D_\mu\theta
)^a)^2 \label{4.26}
\end{equation}
There is one primary constraint,
\begin{equation}
T_1^a= \pi_0^a\approx 0\label{4.27}
\end{equation}
and one secondary constraint,
\begin{equation}
T_2^a = \partial_i\pi_i^a + gf^{abc}\pi_i^bA_i^c -m\pi_\phi^a
-gf^{abc}\pi_\phi^b\phi^c
\label{4.28}
\end{equation}
obtained by time conserving (\ref{4.27}) with the total Hamiltonian,
\begin{equation}
H_T = H_c + \int u^a \pi_0^a \label{4.29}
\end{equation}
where $H_c$ is the canonical Hamiltonian,
\begin{eqnarray}
H_c &=&\frac{1}{2}(\pi_i^a)^2 + \frac{1}{2}(A_i^a)^2 +\frac{1}{4}(F_{ij}^a)^2 +
\frac{1}{2}(\partial_i\phi^a)^2
\frac{1}{2}(\pi_\phi^a)^2+mA_i^a\partial_i\phi^a \nonumber\\
&-&mf^{abc}A_i^b\phi^c\partial_i\phi^a + \frac{1}{2}g^2
f^{abc}f^{ade}A_i^bA_i^d\phi^c\phi^e + A_0^aT_2^a
\label{4.30}
\end{eqnarray}
The algebra of constraints,
\begin{equation}
\{T_1^a(x), T_1^b(y)\} = \{T_1^a(x), T_2^b(y)\} = 0\label{4.31a}
\end{equation}
\begin{equation}
\{T_2^a(x),   T_2^b(y)\}   =  g f^{abc}T_2^c\delta(x-y)   -
gm^2f^{abc}\pi_\theta^c
\delta(x-y) \label{4.31}
\end{equation}
clearly shows that $T_2^a$ is  a  second  class  constraint.  Hence  the  model
(\ref{4.26})  is  not  first  class. In fact, it is an example of a mixed class
system because $T_1^a$ is first class. The relation  in  (\ref{4.31})  was  not
computed  in  \cite{r17}  which  led  its  authors  to incorrectly observe that
(\ref{4.26}) is a first class theory.

\vspace{1cm}
\section {\bf Conclusion}

In the preceeding sections we have made a detailed investigation of the
Batalin-
Tyutin \cite{r8}method of converting second class systems into first class.
Since
the  method  has been developed quite recently it may not be familiar even
among
particle physicists. We have, therefore, reviewed this analysis and  also
shown
the  possibility  of extending it in other directions. In particular, the
method
can be generalised to yield weakly involutive systems  originating  from
second
class  systems.  These  weakly  involutive systems occur naturally in
nonabelian
gauge theories and in gravity. The original paper \cite{r8} is more  suited
for
discussing abelian models.

The systematic application of this approach to specific models-- both abelian
and
nonabelian-- has revealed several interesting features and helps in giving new
insights
than  would  be  possible  by merely discussing the general formalism. The
Proca
model, which is an example of an abelian second class system, is
systematically
converted  into  first  class.  Interestingly,  we  find  that the choice of
the
matrices  $X_{ij}$  (\ref{eq:3.7}),  $\omega^{ij}$  (\ref{eq:3.8}),  which
was
earlier used by us \cite{r18,r19} in other contexts, considerably simplifies
the
algebra.  A  direct  connection  with  the   usual   Lagrangian   embedding
of
St\"uckelberg  \cite{r9}  can  be  made  by  explicitly  evaluating the
momentum
integrals  in  the  phase   space   partition   function   using
Faddeev-Popov
\cite{r14}-like gauges. An exact identification of the extra field introduced
in
our Hamiltonian formalism with the conventional St\"uckelberg  scalar
\cite{r9}
is  possible  provided  one is careful with boundary terms (\ref{3.22}). In
this
connection our analysis may be compared  with  \cite{r12},  which  is  not
only
unsystematic (so may not conform to the strict mathematical rigour of
quantising
first  class  systems  \cite{r3,r4,r5})  but  whose  correspondence   with
the
St\"uckelberg  mechanism is rather circuitous. We find that the present
approach
is more elegant and conceptually clean.

Coming next to the nonabelian  extension  of  the  Proca  model  we  have
first
corrected\footnote{Equation  (\ref{4.8}) given in \cite{r15} is wrong leading
to
an incorrect evaluation of the Dirac brackets involving $A_0^a$. For the
correct
forms, see (\ref{4.n})}  the  familiar
Dirac  analysis  given  in \cite{r15}. Using the generalised canonical
approach,
this model has been coverted into a first class system. Contrary to the
abelian
example,  closed  expressions  for  the  Gauss  constraint  and  the
involutive
Hamiltonian cannot be found. Conceptually it is possible to interpret the
extra
fields  as  the  analogue  of  the  Euler  angles  (regarded as field
variables)
introduced in the nonabelian ($SU(2)$) St\"uckelberg formalism \cite{r16}. It
is
also  possible to obtain first class structure (of constraints, Hamiltonian
etc)
within the Hamiltonian formulation which are  different  from  the
conventional
nonabelian  St\"uckelberg  mechanism  \cite{r16}.  Our  analysis clearly
reveals
that, contrary to existing  claim\footnote{The  algebra  for  $G^a$
immediately
exposes  the  second  class  nature  since  it  does  not  close,  $\{G^a,G^b\}
 =
\epsilon^{abc}G^c + \cdots$.} \cite{r17},  the  ususal  (abelian)
St\"uckelberg
formalism  \cite{r9}  cannot  transform  the  nonabelian  Proca model into
first
class.

We feel that our nonabelian exercise may provide fresh insights in the fields
of
quantum gravity. The point is that it is  possible,  by  suitably  altering
the
generating fuctions, to express the constraint algebra as rank zero, rank one
or
rank two. It may be recalled that a strongly involutive algebra has  rank
zero,
an  algebra  which  closes with structure constants has rank one; and an
algebra
which closes with structure functions of the phase space variables has rank
two.
In  the  case  of quantum gravity Ashtekar \cite{r20} and Witten \cite{r21}
have
given algebras which fall in these clases. It would be tempting to discuss
these
aspects within the generalised canonical framework.

\end{document}